\pdfoutput=1

\documentclass[11pt]{article}

\usepackage[preprint]{acl}

\usepackage{times}
\usepackage{latexsym}
\usepackage[normalem]{ulem}
\useunder{\uline}{\ul}{}

\usepackage[T1]{fontenc}

\usepackage[utf8]{inputenc}

\usepackage{microtype}

\usepackage{inconsolata}

\usepackage{graphicx}

\title{MFMDQwen: Multilingual Financial Misinformation Detection \\Based on Large Language Model}

\author{%
\small
  Zhiwei Liu\textsuperscript{1}\thanks{These authors contributed equally to this work.}\quad
  Yuyan Wang\textsuperscript{1}\footnotemark[1]\quad
  Yuechen Jiang\textsuperscript{1}\quad
  Yupeng Cao\textsuperscript{2}\quad 
  Tianlei Zhu\textsuperscript{3}\quad 
  Xiaorui Guo\textsuperscript{4}\quad \\
\small
  \textbf{Zhiyang Deng}\textsuperscript{2}\quad
  \textbf{Zhiyuan Yao}\textsuperscript{2}\quad 
  \textbf{Xiao-Yang Liu}\textsuperscript{3}\quad
  \textbf{Jimin Huang}\textsuperscript{1,5}\quad 
  \textbf{Sophia Ananiadou}\textsuperscript{1,6} \\ 
\small
    \textsuperscript{1}The University of Manchester \quad 
    \textsuperscript{2}Stevens Institute of Technology \quad 
    \textsuperscript{3}Columbia University \quad \\
\small
    \textsuperscript{4}The University of Edinburgh \quad
    \textsuperscript{5}The Fin AI \quad 
    \textsuperscript{6}ELLIS Manchester \\
\small
\texttt{\{zhiwei.liu,sophia.ananiadou\}@manchester.ac.uk}, ycao33@stevens.edu, tz2617@columbia.edu \\
\small
\texttt{\{yuyan.wang-11,yuechen.jiang,jimin.huang\}@postgrad.manchester.ac.uk} \\
\small
X.Guo-46@sms.ed.ac.uk, zdeng10@stevens.edu, zyao9@stevens.edu, xl2427@columbia.edu
}

\begin{document}
\maketitle
\begin{abstract}
Financial misinformation poses significant threats to financial market stability and individuals’ investment decisions. The multilingual environment and the inherent complexity of financial information present substantial challenges for Multilingual Financial Misinformation Detection (MFMD). Existing LLM-based approaches for financial misinformation detection primarily focus on English and a single financial misinformation detection task, which limits their ability to capture multilingual contexts and complex features. In this paper, we propose MFMDQwen, the first open-source LLM designed for MFMD tasks. Furthermore, we introduce MFMD4Instruction, the first instruction dataset supporting MFMD with LLMs, covering English, Chinese, Greek, and Bengali. We also construct MFMDBench, a benchmark dataset for evaluating the MFMD capabilities of LLMs. Experimental results on MFMDBench demonstrate that our model outperforms existing open-source LLMs. 
The project is available at https://github.com/lzw108/FMD.
\end{abstract}

\section{Introduction}

In the financial domain, the accuracy of information is crucial for investment decisions, financial decision-making, and the stability of financial markets, especially in multilingual environments \cite{rangapur2023investigating,liu2026same}. However, the rapid growth of digital media has exacerbated the spread of financial misinformation. Such misinformation, including deceptive investment advice and misleading statements that distort financial markets, can influence asset prices and broader economic sentiment, thereby posing significant risks \cite{nag2025financial}. The complex characteristics of financial information, such as minor numerical changes, amplified sentiment, and reversed causality, present substantial challenges for automated detection \cite{jiang2026all}. Time-consuming manual inspection is clearly unsuitable for detecting the rapidly evolving and large volume of financial misinformation, and the development of large language models (LLMs) has made automated detection methods possible.

In recent years, LLMs with massive numbers of parameters have emerged as a novel approach to tackling various problems in the financial domain, achieving remarkable results \cite{li2023large}. This includes applications in financial misinformation detection, such as FMDLlama \cite{liu2025fmdllama}, and the creation of financial misinformation datasets like FinFact \cite{rangapur2023finfact}, FinDver \cite{zhao2024findver}, and RFCBench \cite{jiang2026all}, as well as studies on bias \cite{liu2026same}. Nevertheless, most research still focuses on English data, overlooking the more complex and higher-risk problem of multilingual financial misinformation. 

To tackle these issues, we construct the first instructing-tuning dataset for multilingual financial misinformation detection (MFMD4Instruction) to support LLMs' supervised fine-tuning (SFT). Based on MFMD4Instruction, we developed the first open-source multilingual financial misinformation model through SFT to support the MFMD task across multiple languages, including English, Chinese, Greek, and Bengali. To evaluate the financial misinformation verification ability of LLMs, we also built a benchmark for the detection of multilingual financial misinformation (MFMDBench). We evaluated MFMDQwen and numerous baselines on MFMDBench, and the results demonstrate that MFMDQwen achieved the best overall performance across all nine datasets.

Our main contributions are as follows:

(1) We construct MFMD4Instruction, the first multilingual financial misinformation dataset for SFT of LLMs.

(2) We develop MFMDQwen, the first open-source LLM for multilingual financial misinformation detection.

(3) We build MFMDBench, the first benchmark to evaluate the verification ability of multilingual financial misinformation of LLMs, including 9 tasks and covering 4 languages. The results on MFMDBench demonstrate that our model outperforms other open-source LLMs.

\section{Related work}

\subsection{Financial Misinformation Detection}

In finance, where accurate information is essential for market stability, and trust, digital media has accelerated the spread of misinformation \cite{rangapur2023investigating}. Recent studies have explored automated financial misinformation detection with LLMs. \citet{rangapur2023finfact} propose the dataset for financial fact checking and explanation generation, and evaluate the ability of several LLMs. FMDLlama applies instruction tuning to adapt LLMs to this task \cite{liu2025fmdllama}, while the FinFact workshop brought together diverse approaches on a shared dataset \cite{liu2025finnlp}. Other work addresses financial data scarcity through general-domain augmentation, evidence generation, and few-shot retrieval \cite{lee2025dunamu}, or extends detection to multimodal settings by combining text with image-derived descriptions \cite{luo2025fmd}. Beyond detection, \citet{cao2025capybara} improve financial reasoning by combining retrieved evidence with a Financial Chain-of-Thought framework, and FinDVer provides a benchmark for explainable claim verification on long, hybrid financial documents, showing that even GPT-4o trails human experts \cite{zhao2024findver}. More recently, MFMDBench~\cite{liu2026same} introduced a multilingual benchmark for financial misinformation detection across diverse cultural contexts. \textsc{RFC-Bench}~\cite{jiang2026all} further investigates professional financial misinformation detection in a reference-free setting. Both benchmarks reveal that current LLMs still exhibit substantial limitations when handling financial misinformation detection in complex and professional scenarios.

\subsection{Open-sourced Large Language Models}

LLMs have been widely applied across various fields and have achieved good results. For example, the ChatGPT series \cite{openai_gpt41}, Deepseek \cite{deepseek_models}, and LLama \cite{llama31_techreport}. There is also a lot of research dedicated to open-sourcing domain-specific LLMs, such as LLMs for finance \cite{wu2023bloomberggpt,xie2023pixiu}, health \cite{chen2024huatuogpt,xiao2025mentrasuite}, misinformation detection \cite{liu2025conspemollm,liu2026raar}, and so on. These open-source models have facilitated in-depth research in their respective domains. For financial misinformation, the existing FMDllama \cite{liu2025fmdllama} only supports English and cannot be applied to multilingual financial misinformation detection. Therefore, this paper develops MFMDQwen for multilingual misinformation detection.

\section{Methods}

\begin{figure}[htb]
\centering
  \includegraphics[width=0.8\columnwidth]{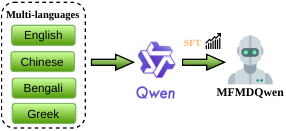}
  \caption{The architecture of MFMDQwen.}
  \label{fig:mainmethod}
\end{figure}

\subsection{Task formalization}

We formulate financial misinformation detection as a generative task, leveraging a generative model as the foundation. Specifically, we adopt an autoregressive language model $\mathbf{P}_{\phi}(y \mid x)$ parameterized by pre-trained weights $\phi$. This model is capable of simultaneously handling multiple multilingual financial misinformation detection tasks. Each task $t$ is defined as a set of context-target pairs:

\[
D_t = \{(q_i^t, r_i^t)\}_{i=1}^{N_t},
\]
where: $q_i^t$ is a context token sequence that includes the task description, input text, and query; $r_i^t$ is a target token sequence that represents the final answer.

To enable multi-task learning, all task datasets are merged into a single dataset. The model is then trained to maximize the conditional likelihood of the target sequences given the contexts:
\[
\max_{\phi} \sum_{t} \sum_{i=1}^{N_t} \log \mathbf{P}_{\phi}(r_i^t \mid q_i^t).
\]

Through this setup, the model learns to generate final answers, improving prediction accuracy via SFT.

\subsection{MFMD4Instruction}

\subsubsection{Raw Data Collection}

Based on MFMDScen \cite{liu2026same} and RFC \cite{jiang2026all}, we compile a multilingual financial misinformation corpus from nine existing sources spanning four languages: English, Chinese, Bengali, and Greek. These datasets cover a range of related tasks, including claim verification, fact-checking, misinformation detection, and social media manipulation detection.

\paragraph{English.}
\textbf{FinDVer}~\cite{zhao2024findver} contains 700 financial claim verification instances derived from financial reports. Each instance consists of a financial statement paired with relevant evidence paragraphs from the corresponding report. The dataset is balanced, with 350 \textit{entailed} and 350 \textit{refuted} samples. \textsc{\textbf{RFC-Bench}}~\cite{jiang2026all} is a paragraph-level benchmark for reference-free counterfactual financial misinformation detection. It contains 1,826 original–perturbed paragraph pairs derived from real Yahoo Finance articles covering 223 U.S. stocks. Perturbations are generated via GPT-4.1 under four manipulation types (Directional Flipping, Numerical Perturbation, Sentiment Amplification, and Causal Distortion) and validated through multi-stage expert review and dual-annotator evaluation to ensure label reliability.

\paragraph{Chinese.}
\textbf{CHEF} \cite{hu2022chef} is a Chinese fact-checking dataset with 1,188 samples, each comprising a claim and associated evidence texts. Its labels are \textit{supported}, \textit{refuted}, and \textit{not enough information}.  
\textbf{MDFEND} \cite{nan2021mdfend} contains 1,321 Weibo posts labeled as either \textit{real} (959 samples) or \textit{fake} (362 samples).

\paragraph{Bengali.}
\textbf{Bengali Manipulation}~\cite{kamruzzaman2023banmani} (\textbf{BanMANI}) includes 101 samples for manipulated social media detection. Each sample pairs an original news article with a related social media post and is labeled as \textit{MANI} (52 samples) or \textit{NO\_MANI} (49 samples).

\paragraph{Multilingual.}
\textbf{Global4Languages}~\cite{liu2026same} provides aligned claim verification samples in English, Chinese, Bengali, and Greek, with 144 instances per language. Each sample contains a claim and its scenario context, labeled as \textit{true} (23 samples) or \textit{false} (121 samples).

\begin{table}[]
\footnotesize
\centering
\resizebox{0.48\textwidth}{!}{
\begin{tabular}{lcccc}
\hline
Dataset   & Language & MFDM4Instruction & MFMDBench & Total \\ \hline
FinDVer   & EN   & 554              & 140       & 694   \\
CHEF      & ZH   & 894              & 238       & 1132  \\
MDFEND    & ZH   & 1044             & 265       & 1309  \\
Bengali   & BN   & 80               & 21        & 101   \\
Global-EN & EN   & 115              & 29        & 144   \\
Global-ZH & ZH   & 115              & 29        & 144   \\
Global-BN & BN   & 115              & 29        & 144   \\
Global-GR & GR   & 115              & 29        & 144   \\
RFC       & EN   & 1805             & 1652      & 3457  \\
Total     & -    & 4837             & 2432      & 7269  \\ \hline
\end{tabular}
}
\caption{Statistics of the datasets. EN: English,
ZH: Chinese, BN: Bengali, GR: Greek.}
\label{tab:rawdata}
\end{table}

\subsubsection{Construction of Base Data for Complex Reasoning}
\label{sec:basedata}

To prepare the raw datasets for complex reasoning path construction, we convert each dataset into a unified instruction-tuning format. Each sample is transformed into a structured record containing: (1) a \textbf{task description} specifying the detection task and expected output format, (2) the \textbf{input fields} consisting of the claim or content together with any supporting evidence or contextual information. All templates can be found at Appendix \ref{app:template}.





\subsection{MFMDQwen}

\begin{table*}[]
\resizebox{1\textwidth}{!}{
\begin{tabular}{lccccccccccccccccccc}
\hline
Models       & \multicolumn{2}{c}{GlobalGr}    & \multicolumn{2}{c}{GlobalBe}    & \multicolumn{2}{c}{GlobalCh}    & \multicolumn{2}{c}{GlobalEn}    & \multicolumn{2}{c}{CHEF}        & \multicolumn{2}{c}{MDFEND}      & \multicolumn{2}{c}{Bengali}     & \multicolumn{2}{c}{FinDver}     & \multicolumn{2}{c}{RFC}         & Ave.           \\
             & ACC            & F1             & ACC            & F1             & ACC            & F1             & ACC            & F1             & ACC            & F1             & ACC            & F1             & ACC            & F1             & ACC            & F1             & ACC            & F1             & F1             \\ \hline
Qwen3-8b-R   & 0.862          & 0.710          & 0.655          & 0.396          & \textbf{0.862} & \textbf{0.710} & 0.793          & 0.638          & 0.420          & 0.207          & 0.668          & 0.652          & 0.905          & 0.905          & 0.807          & 0.806          & 0.668          & 0.652          & 0.631          \\
Qwen3-14b-R  & 0.897          & 0.756          & 0.759          & 0.607          & 0.862          & 0.628          & \textbf{0.897} & \textbf{0.756} & 0.429          & 0.272          & 0.555          & 0.553          & 0.762          & 0.760          & 0.814          & 0.814          & 0.555          & 0.553          & 0.633          \\
Qwen3-32b-R  & 0.793          & 0.638          & 0.793          & 0.685          & \textbf{0.862} & \textbf{0.710} & 0.828          & 0.671          & 0.424          & 0.275          & 0.653          & 0.639          & 0.952          & 0.952          & \textbf{0.857} & \textbf{0.857} & 0.653          & 0.639          & 0.674          \\
Qwen3-8b     & 0.862          & 0.710          & 0.828          & 0.671          & 0.759          & 0.431          & 0.862          & 0.628          & 0.462          & 0.320          & 0.608          & 0.604          & 0.952          & 0.952          & 0.750          & 0.750          & 0.608          & 0.604          & 0.630          \\
Qwen3-14b    & 0.759          & 0.431          & 0.793          & 0.685          & 0.828          & 0.671          & 0.828          & 0.671          & 0.429          & 0.272          & 0.623          & 0.617          & 0.762          & 0.741          & 0.793          & 0.793          & 0.623          & 0.617          & 0.611          \\
Qwen3-32b    & 0.862          & 0.628          & 0.828          & 0.671          & 0.828          & 0.671          & 0.759          & 0.607          & 0.458          & 0.310          & 0.709          & 0.689          & 0.952          & 0.952          & 0.793          & 0.793          & 0.709          & 0.689          & 0.668          \\
Qwen2.5-72b  & 0.862          & 0.710          & 0.793          & 0.565          & 0.759          & 0.540          & 0.828          & 0.671          & 0.521          & 0.320          & 0.857          & 0.819          & 0.952          & 0.952          & 0.843          & 0.841          & 0.551          & 0.543          & 0.662          \\
Llama3.1-8b  & 0.621          & 0.466          & 0.207          & 0.133          & 0.690          & 0.427          & 0.759          & 0.463          & 0.500          & 0.203          & 0.664          & 0.429          & 0.429          & 0.276          & 0.721          & 0.484          & 0.664          & 0.429          & 0.368          \\
Llama3.3-70b & \textbf{0.931} & \textbf{0.879} & 0.759          & 0.470          & 0.759          & 0.653          & 0.793          & 0.638          & 0.471          & 0.279          & 0.815          & 0.763          & \textbf{1.000} & \textbf{1.000} & 0.821          & 0.818          & 0.529          & 0.466          & 0.663          \\
FMDLlama     & 0.759          & 0.431          & 0.069          & 0.051          & 0.828          & 0.594          & 0.793          & 0.383          & 0.580          & 0.193          & 0.272          & 0.143          & 0.429          & 0.299          & 0.614          & 0.402          & 0.272          & 0.143          & 0.293          \\
MFMDQwen         & 0.862          & 0.758          & \textbf{0.897} & \textbf{0.803} & \textbf{0.862} & \textbf{0.710} & \textbf{0.897} & \textbf{0.756} & \textbf{0.849} & \textbf{0.843} & \textbf{0.932} & \textbf{0.913} & 0.952          & 0.952          & 0.700          & 0.673          & \textbf{0.954} & \textbf{0.954} & \textbf{0.818} \\ \hline
\end{tabular}
}
\caption{Results on MFMDBench.}
\label{tab:results}
\end{table*}

\begin{table}[]

\end{table}

Figure \ref{fig:mainmethod} presents the overview of MFMD-R. We first construct the reasoning paths (MFMD4Instruction). We then built MFMD-R based on Qwen-3-8B \cite{qwen3_techreport} using the MFMD4Instruction dataset. The model is trained in two stages: SFT followed by RL. During SFT, the learning rate is set to $1\times10^{-5}$ with a warmup ratio of 0.1. Training runs for 2 epochs with a batch size of 128. DeepSpeed ZeRO-3 optimization is used with CPU parameter offloading to reduce GPU memory usage. The maximum input sequence length is 24k tokens and the maximum output length is 8k tokens. Full-parameter training is conducted on four NVIDIA L40s GPUs (48 GB).

\section{Experiments}

\subsection{Baseline models}


Our extensive evaluation of open-source and proprietary LLMs included the following reasoning-focused models: Qwen3 reasoning variants (8B-R, 14B-R, and 32B-R), Qwen3 no-reasoning models (8B, 14B, and 32B) \cite{qwen3_techreport}, Qwen2.5-72B-Instruct \cite{qwen25_official}, Llama-3.1-8B-Instruct and Llama-3.3-70B-Instruct \cite{llama31_techreport}. We also compare domain-specific financial misinformation LLMs (i.e. FMDLlama \cite{liu2025fmdllama})

\subsection{Evaluation methods}

We uses metrics such as Accuracy, Macro-F1 for misinformation detection evaluation.

\subsection{Results}

Table \ref{tab:results} presents the experimental results on the MFMDBench dataset. In the following analysis, we primarily focus on the F1 score. As shown in the table, MFMDQwen attains top scores on 6 of 9 benchmarks, including models of comparable size as well as larger LLMs with 14B, 32B, and 72B parameters. These results indicate that MFMDQwen demonstrates clear advantages in multilingual tasks and validate the effectiveness of SFT training. This strategy significantly enhances the multilingual representation capabilities of LLMs. Additional confusion matrix visualizations are provided in Appendix~\ref{app:confusion_matrices}.

It is also worth noting that some LLMs occasionally produce uncertain or irrelevant responses when answering questions, which partially explains the relatively lower F1 scores. This phenomenon is particularly evident in Llama3.1-8B and FMDLlama, which frequently generate responses indicating that they are unable to provide an answer. This behavior may be related to their relatively strict safety or protection mechanisms. Additionally, models with reasoning capabilities do not always outperform non-reasoning models of the same size, as the additional reasoning steps may lead to overthinking and ultimately produce incorrect answers.

Furthermore, performance differences across languages reveal additional insights into model behavior. MFMDQwen shows the most consistent gains on Chinese datasets (CHEF and MDFEND), where it significantly outperforms all baselines. This suggests that the proposed training strategy is particularly effective in handling noisy, context-dependent misinformation commonly found in social media environments. In contrast, on English datasets, the performance gap between MFMDQwen and strong LLM baselines is relatively smaller. For instance, on FinDVer, some larger models achieve comparable or even better results, indicating that structured evidence-based judgment remains a strength of general-purpose LLMs.

For low-resource languages such as Bengali, most models achieve near-saturated performance, which can be attributed to the limited dataset size and reduced task complexity. On the GlobalGr dataset, MFMDQwen ranks second, slightly below Llama3.3-70B, while still outperforming most other baselines. Notably, it achieves the best performance on GlobalBe. These results further validate the effectiveness of the proposed fine-tuning strategy, showing that it enables strong cross-lingual generalization and competitive performance even against substantially larger models.

Overall, MFMDQwen demonstrates superior cross-lingual stability compared to both general-purpose LLMs and the domain-specific FMDLlama model. While FMDLlama is tailored for financial misinformation detection, its performance varies significantly across languages, indicating limited multilingual generalization. In contrast, MFMDQwen benefits from supervised fine-tuning on multilingual datasets, enabling it to capture more robust and language-invariant misinformation patterns across diverse linguistic settings.

\section{Conclusion}

In this paper, we propose MFMDQwen, the first open-sourced LLM for Multilingual Financial Misinformation Detection (MFMD). We also construct a multi-task multilingual financial misinformation dataset (MFMD4Instruction) and an MFMD evaluation benchmark (MFMDBench). We conduct a comprehensive evaluation of MFMDQwen and a variety of LLMs on the MFMDBench benchmark. The results show that MFMDQwen achieves the best performance on 6 out of 9 datasets, demonstrating its strong capability in MFMD tasks. 

In future work, we plan to expand MFMD4Instruction and MFMDBench to include more languages, integrate additional financial-related tasks, and further improve the overall performance of MFMDQwen.

\section{Limitations}

Due to restricted computational resources and cost, we only carried out instruction-tuning/evaluation of multilingual financial misinformation detection tasks using 8b/14b/32b LLMs. As such, we have not considered the impact of using larger models on the MFMDBench tasks.




\bibliography{acl_latex}

\appendix

\section{Prompt templates \label{app:template}}

Based on MFMDScen \cite{liu2026same} and RFC \cite{jiang2026all}, we define the following prompts as templates:

\begin{center}
\footnotesize
\fcolorbox{black}{gray!10}{
\begin{minipage}{0.45\textwidth}
\footnotesize
\textbf{Prompt template for Global4Languages:}  \\

\textbf{Task Description:} Determine whether the claim is 'True' or 'False'.

\textbf{Claim:} \textit{[Claim]}

\end{minipage}
}
\end{center}

\begin{center}
\footnotesize
\fcolorbox{black}{gray!10}{
\begin{minipage}{0.45\textwidth}
\footnotesize
\textbf{Prompt template for FinDVer:}  \\

\textbf{Task Description:} Assess the truthfulness of the given statement by determining whether it is entailed or refuted based on the provided financial document. Output the entailment label (‘entailed’ or ‘refuted’) of the claim.

\textbf{Claim:} \textit{[Claim]}

\textbf{Relevant Financial Report:} \textit{[Document]}

\end{minipage}
}
\end{center}

\begin{center}
\footnotesize
\fcolorbox{black}{gray!10}{
\begin{minipage}{0.45\textwidth}
\footnotesize
\textbf{Prompt template for MDFEND:}  \\

\textbf{Task Description:} Determine whether the following content is 'real' or 'false'.

\textbf{Content:} \textit{[Content]}

\end{minipage}
}
\end{center}

\begin{center}
\footnotesize
\fcolorbox{black}{gray!10}{
\begin{minipage}{0.45\textwidth}
\footnotesize
\textbf{Prompt template for CHEF:}  \\

\textbf{Task Description:} Label each claim based on the evidence provided. Choose one of the following three labels: Supported, which means there is sufficient evidence showing the claim is supported; Refuted, which means there is sufficient evidence showing the claim is refuted; Not enough information, which means the evidence is insufficient to determine whether the claim is supported or refuted.

\textbf{Claim:} \textit{[Claim]}

\textbf{Evidence:} \textit{[Evidence]}

\end{minipage}
}
\end{center}

\begin{center}
\footnotesize
\fcolorbox{black}{gray!10}{
\begin{minipage}{0.45\textwidth}
\footnotesize
\textbf{Prompt template for BanMANI:}  \\

\textbf{Task Description:} Determine whether the social media post is manipulated or not manipulated based on the original news. Output `MANI' in case the post is manipulated from the original news article,  or output `NO\_MANI' otherwise.

\textbf{Original News:} \textit{[Original News]}

\textbf{Social Media Post:} \textit{[Social Media Post]}

\end{minipage}
}
\end{center}

\begin{center}
\footnotesize
\fcolorbox{black}{gray!10}{
\begin{minipage}{0.45\textwidth}
\footnotesize
\textbf{Prompt template for RFC:}  \\

\textbf{Task Description:} You are a financial misinformation detector. Please check whether the following information is true or false and output the answer [true/false].

\textbf{News:} \textit{[News]}

\end{minipage}
}
\end{center}

\section{Confusion Matrix Visualizations}
\label{app:confusion_matrices}

\begin{figure}[ht]
    \centering
    \includegraphics[width=\linewidth]{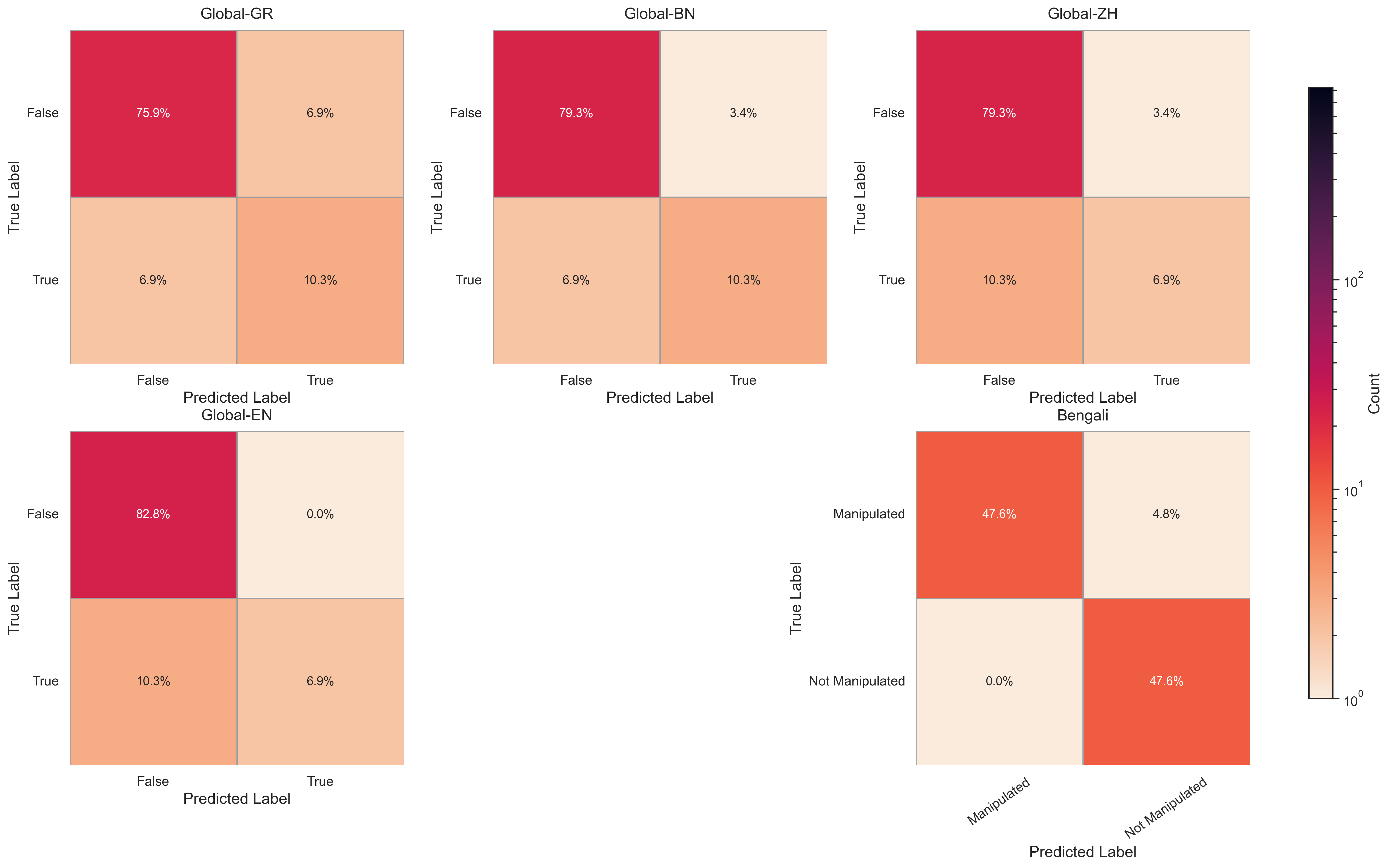}
    \caption{
    Confusion matrices for five binary classification datasets (GlobalGr, GlobalBe, GlobalCh, GlobalEn, and Bengali).
    We apply logarithmic normalization to improve visual contrast across datasets with different scales.
    Rows correspond to ground-truth labels and columns correspond to model predictions.
    }
    \label{fig:cm_group1}
\end{figure}

\begin{figure}[ht]
    \centering
    \includegraphics[width=\linewidth]{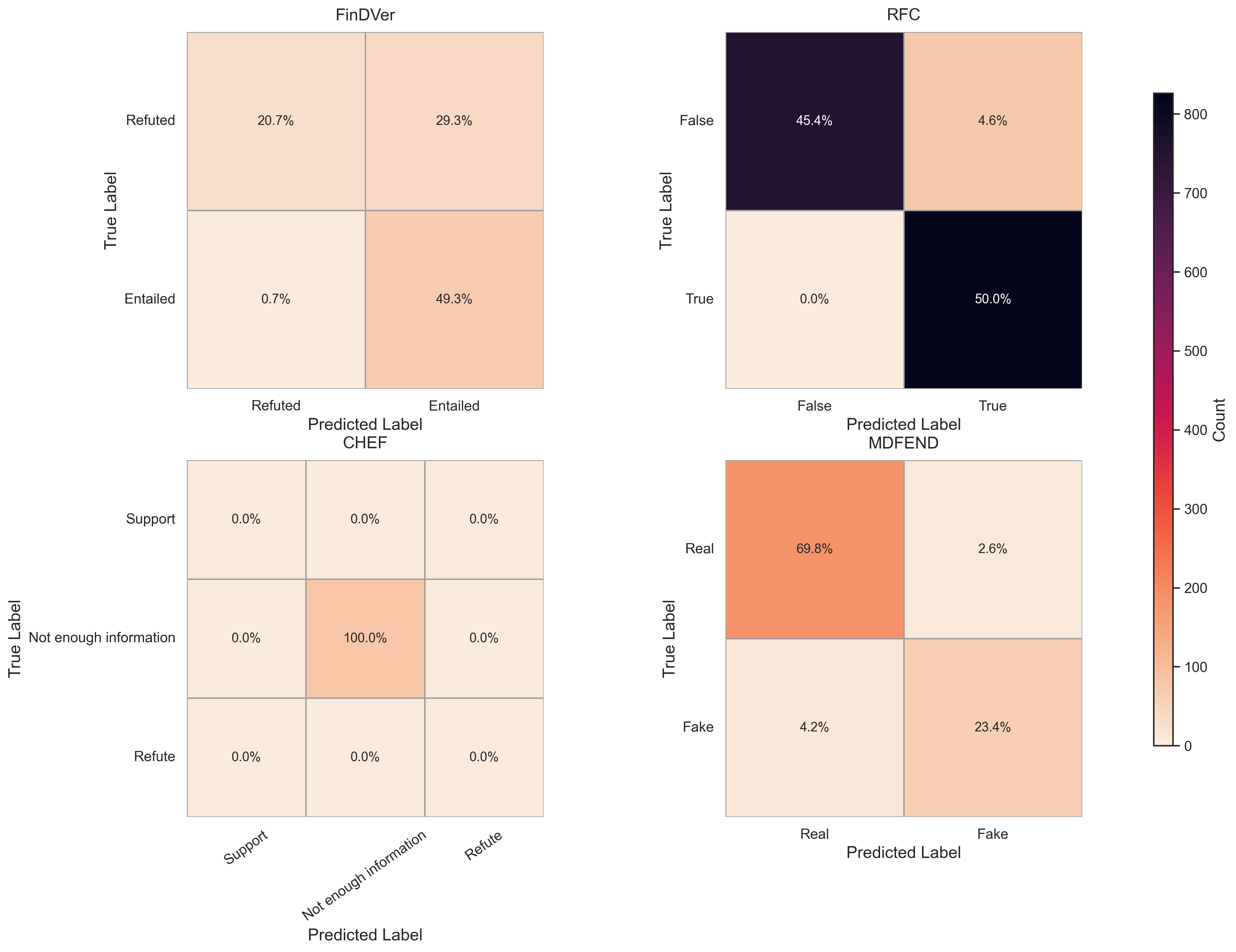}
    \caption{
    Confusion matrices for four datasets with heterogeneous label spaces (Chef, MDFEND, FindVer, and RFC).
    A shared linear normalization is used to preserve the relative magnitude of counts across datasets.
    }
    \label{fig:cm_group2}
\end{figure}

\newpage
\section{Error Case Study}
\label{app:error_cases}

We manually inspected representative prediction errors across datasets and identified five recurring error types. These errors suggest that the model does not fail uniformly; rather, its mistakes stem from distinct weaknesses in polarity control, numerical reasoning, evidence grounding, and fine-grained semantic comparison.

\subsection{Error Type 1: Label Polarity Reversal and Prior-Driven Guessing}

A common failure mode in multilingual true/false datasets is label polarity reversal, in which the model predicts the opposite label despite the dataset's simple binary format. In several cases, the model appears to rely on surface plausibility or world-knowledge priors rather than on the benchmark label.

\paragraph{Case Study.}
In \textbf{Global-EN}, the model predicts \textit{False} for the claim: \textit{``Amazon solicited donations from the public to pay sick leave to contractors and seasonal workers during the COVID-19 pandemic.''} The gold label is \textit{True}. Similar polarity reversals also appear in \textbf{Global-GR}, \textbf{Global-BN}, and \textbf{Global-ZH} for translated variants of the same claim. This cross-lingual consistency suggests that the error is not solely due to translation difficulty but to a systematic tendency to default to a plausible-sounding negative judgment.

\subsection{Error Type 2: Numerical Reasoning and Arithmetic Errors}

The model frequently fails on claims that require explicit calculation, especially percentage change, percentage decrease, or net value derived from financial figures. These errors indicate weak arithmetic grounding even when all necessary numbers are explicitly present in the document.

\paragraph{Case Study.}
In \textbf{FinDVer}, the model predicts \textit{Entailed} for the claim: \textit{``The percentage increase in the working capital deficit from March 31, 2023, to December 31, 2023, is approximately 38.33\%.''} The gold label is \textit{Refuted}. The report states that the working capital deficit increased from \$84{,}255 to \$116{,}603. This requires an explicit percentage-change calculation, and the model incorrectly accepts the claimed value rather than verifying it against the numbers.

\subsection{Error Type 3: Table Parsing and Financial Statement Grounding Errors}

Another major source of errors is incorrect grounding in semi-structured financial tables. The model often extracts the wrong row, wrong column, or wrong derived quantity, especially when multiple time periods and measures are presented together.

\paragraph{Case Study.}
In \textbf{FinDVer}, the model predicts \textit{Entailed} for the claim: \textit{``The total operating margin percentage for Kennametal over the six months ended December 31, 2023, is 7.4\%.''} The gold label is \textit{Refuted}. The evidence is presented in a multi-column table with total sales and total operating income for different periods. Correct judgment requires selecting the correct six-month values and computing the margin from them. The model appears to accept the stated percentage without robustly grounding it in the table.

\subsection{Error Type 4: Partial Evidence Matching and Missing Critical Detail}

The model also fails when a claim is partially supported by the evidence but includes one critical, incorrect detail. In such cases, it appears to match the generally relevant topic while overlooking the exact attribute that determines the label.

\paragraph{Case Study.}
In \textbf{FinDVer}, the model predicts \textit{Entailed} for the claim: \textit{``In 2023, AMN Healthcare saw 54\% of its consolidated revenue flow through Managed Services Programs, while Kaiser Foundation Hospitals accounted for approximately 27\% of its total consolidated revenue.''} The gold label is \textit{Refuted}. The first part of the claim is supported by the report, but the second part is not: the report states that Kaiser accounted for approximately \textbf{17\%}, not 27\%. The model likely overrelied on partial support from the first clause and failed to carefully verify the second clause.

\subsection{Error Type 5: Manipulation Detection Failure under Subtle Lexical Mismatch}

For manipulation detection, the model struggles when a social media post preserves the original news's general topic but alters a key factual detail. These errors show weak sensitivity to subtle semantic distortion rather than complete topic mismatch.

\paragraph{Case Study.}
In \textbf{Bengali}, the model predicts \textit{Not Manipulated} for the post: \textit{``City Bank-American Express launches the first airport lounge.''} The gold label is \textit{Manipulated}. The original news states that the newly launched lounge is their \textbf{second international lounge}, while also mentioning that the company had launched the country's first airport lounge earlier. The manipulated post changes the event-specific fact from \textit{second} to \textit{first}. The model appears to match the overall topic correctly but misses the crucial ordinal inconsistency.

\subsection{Discussion}

Overall, the observed errors indicate that the model's weaknesses are not limited to a single dataset or language. Instead, they cluster around a small set of recurring reasoning failures: polarity instability in binary classification, weak arithmetic verification, fragile grounding in financial tables, over-acceptance under partial evidence overlap, and insufficient sensitivity to subtle factual distortions. These findings suggest that future improvements should focus not only on general instruction following but also on explicit verification mechanisms for numbers, table structure, and claim-level fact alignment.

\end{document}